\documentclass[prb,twocolumn,nolongbibliography,superscriptaddress]{revtex4-2}

\usepackage[colorlinks,bookmarks=true,citecolor=blue,linkcolor=red,urlcolor=blue,citecolor=blue]{hyperref}

\usepackage{amsmath}
\usepackage{amssymb}
\usepackage{mathtools}
\usepackage{amsfonts}
\usepackage{bbm}
\usepackage{braket}
\usepackage{xcolor}
\usepackage{appendix}
\usepackage{enumitem}
\usepackage{graphicx}
\usepackage{float}
\usepackage{tikz}
\usepackage{algorithm}
\usepackage{algpseudocode}
\usepackage{bbm}

\usepackage[enableskew,vcentermath]{youngtab}

\usepackage{subfigure}
\usepackage{color}
\usepackage{nicematrix}

\newcommand{\SM}{\fontsize{2pt}{\baselineskip}\selectfont}
\newcommand{\sm}{\fontsize{6pt}{\baselineskip}\selectfont}
\newcommand{\LG}{\fontsize{14pt}{\baselineskip}\selectfont}

\begin{document}

\title{Bridging conformal field theory and parton approaches to $\mathrm{SU}(n)_k$ chiral spin liquids}

\author{Tong Liu}
\affiliation{Institute of Physics, Chinese Academy of Sciences, Beijing 100190, China}
\affiliation{School of Physical Sciences, University of Chinese Academy of Sciences, Beijing 100049, China}

\author{Ying-Hai Wu}
\affiliation{School of Physics and Wuhan National High Magnetic Field Center, Huazhong University of Science and Technology, Wuhan 430074, China}

\author{Hong-Hao Tu}
\email{h.tu@lmu.de}
\affiliation{Faculty of Physics and Arnold Sommerfeld Center for Theoretical Physics, Ludwig-Maximilians-Universit\"at M\"unchen, 80333 Munich, Germany}

\author{Tao Xiang}
\email{txiang@iphy.ac.cn}
\affiliation{Institute of Physics, Chinese Academy of Sciences, Beijing 100190, China}
\affiliation{School of Physical Sciences, University of Chinese Academy of Sciences, Beijing 100049, China}
\affiliation{Beijing Academy of Quantum Information Sciences, Beijing 100190, China}

\begin{abstract}
We employ the $\mathrm{SU}(n)_k$ Wess-Zumino-Witten (WZW) model in conformal field theory to construct lattice wave functions in both one and two dimensions. The spins on all lattice sites are chosen to transform under the $\mathrm{SU}(n)$ irreducible representation with a single row and $k$ boxes in the Young tableau. It is demonstrated that the wave functions can be reinterpreted as parton states, which enables efficient conversion to matrix product states such that many physical properties can be evaluated directly. In one dimension, these wave functions describe critical spin chains whose universality classes are in one-to-one correspondence with the WZW models used in the construction. In two dimensions, our constructions yield model wave functions for chiral spin liquids, and we show how to find all topological sectors of them in a systematic way. Using the null vectors of Kac-Moody algebras, parent Hamiltonians of the $\mathrm{SU}(3)_k$ series are derived. The $\mathrm{SU}(3)_k$ chiral spin liquids are lattice analogs of non-Abelian spin-singlet fractional quantum Hall states, and the $k=2$ member hosts Fibonacci anyons. 
\end{abstract}

\maketitle

\section{Introduction}

In a quantum many-body system, the wave function should be intact or acquire a minus sign when two particles are exchanged, depending on whether they are bosons or fermions. If this exchange is elevated to an adiabatic process called braiding, nontrivial phases may be generated when two point-like objects are operated in two dimensions~\cite{Leinaas1977}. These exotic objects were named as anyons and connected with representations of the braid group~\cite{Wilczek1982,WuYS1984a}. The remarkable discovery of fractional quantum Hall (FQH) effect in two-dimensional electron gases brings anyons to the forefront of condensed matter physics in the past four decades~\cite{Tsui1982,Laughlin1983,Arovas1984}. Another interesting class of phases that supports anyons is quantum spin liquids~\cite{ZhouY2017}. In some cases, intimate connections between FQH states and spin liquids can be established~\cite{Kalmeyer1987}. The existence of anyons leads to topology dependent ground-state degeneracy on closed manifolds and hence the concept of topological order was introduced~\cite{WenXG1989}. It was further proposed that some FQH states support non-Abelian anyons~\cite{Moore1991,WenXG1991a}, which are related to higher-dimensional representations of the braid group. Braiding two non-Abelian anyons executes a unitary transformation on the system, and this may be harnessed to perform topological quantum computation~\cite{RMP_TQC}.

Theoretical understanding of topological orders relies heavily on the construction of model wave functions. For example, Laughlin provided an insightful and elegant explanation of the $1/3$ FQH state in this manner~\cite{Laughlin1983}. More generally, two approaches have been demonstrated as fruitful by extensive investigations over the past years. The first one utilizes multipoint correlation functions in certain conformal field theories (CFTs)~\cite{Moore1991}. The bulk wave function for a system is obtained using the CFT that describes its edge physics, providing a vivid illustration of bulk-edge correspondence. In this framework, the Laughlin wave function is viewed as the correlator of vertex operators of chiral bosons $V(z)=e^{i\sqrt m \phi(z)}$. The Moore-Read Pfaffian wave function was obtained using the operator $V(z)=\chi(z)e^{i\sqrt m \phi(z)}$ with an extra Majorana field~\cite{Moore1991}, and its elementary charged excitations are Ising anyons. The second one treats each physical degrees of freedom as a combination of multiple partons that are glued together by emergent gauge fields~\cite{WenXG1991a}. A strongly correlated state of the system is obtained by placing each species of partons in a weakly correlated mean-field state.

The CFT correlator approach has also been extended to lattice systems~\cite{Cirac2010,Nielsen2011,Nielsen2012,TuHH2013b,Nielsen2014,TuHH2014b,TuHH2014a,Bondesan2014,TuHH2015,Deshpande2016,Hackenbroich2017,ZhangHC2021,ZhangHC2022} to construct a variety of chiral spin liquids using the $\mathrm{SU}(2)_1$~\cite{Nielsen2011,Nielsen2012}, $\mathrm{SU}(2)_2$~\cite{Nielsen2011}, $\mathrm{SO}(n)_1$~\cite{TuHH2013b} and $\mathrm{SU}(n)_1$~\cite{TuHH2014a,Bondesan2014} Wess-Zumino-Witten (WZW) models~\cite{Francesco-Book}. In many cases, parent Hamiltonians for which these spin liquids are exact ground states can be 
derived using the null vectors of the corresponding Kac-Moody algebra, which typically contain long-ranged and multi-spin interaction terms. One may truncate the interaction range and tune the coefficients of the remaining terms. Numerical calculations using such modified Hamiltonians have gathered evidence for chiral spin liquids~\cite{Nielsen2013,Glasser2015,WuYH2016a,ChenJY2018,Nandy2019,ChenJY2021,Jaworowski2022,HuangYX2022,XuY2023}. All cases listed above are WZW models with free field representations, whose correlation functions (ground-state wave functions in the spin models) can be easily evaluated. The cases of interacting CFTs are more difficult and much less explored~\cite{Quella2020}.

This paper reports our investigations on lattice wave functions of spin systems constructed from interacting $\mathrm{SU}(n)_k$ WZW models in both one- and two-dimensional cases (some of these models have been studied in Refs.~\cite{Greiter2009,Michaud2013,Meng2015,Nataf2016a,Nataf2016b,Lecheminant2017,Fuji2017,ChenJH2017,LiuZX2018,YaoY2019,Ferraz2019,Iadecola2019,Chepiga2022,Xavier2023,Chepiga2024} using different analytical or numerical methods). It is required that all spins transform under the $\mathrm{SU}(n)$ irreducible representation with a single row and $k$ boxes in the Young tableau. A fundamental breakthrough is that these wave functions can be recast as projected Fermi seas (one subclass of the parton states). In the context of FQH physics, the connection between CFT and parton approaches has been studied~\cite{Hendersen2024,Anand2022,Fuchs1994}, but an analogous relationship for chiral spin liquids in lattice systems remains largely unexplored. At the conceptual level, this discovery unifies the CFT and parton approaches for studying topological order, which invites further exploration of deeper implications. At the practical level, the parton representations can be expressed as matrix product states (MPSs) using the method developed by the present authors~\cite{LiuT2025a}, which allows for efficient computation of various physical properties using standard techniques. We confirm that the one-dimensional (1D) wave functions describe critical spin chains. For two-dimensional (2D) systems, our main focus is the $\mathrm{SU}(3)_k$ series. In the quantum Hall context, such CFTs were employed to construct non-Abelian spin-singlet FQH states~\cite{Ardonne1999,Ardonne2001}. The $\mathrm{SU}(3)_2$ chiral spin liquid supports Fibonacci anyons, which are highly desirable for their capability of universal topological quantum computation~\cite{RMP_TQC}. Parent Hamiltonians for these lattice wave functions are also derived.

The rest of this paper is organized as follows. In Sec.~\ref{sec:MWF}, we show explicitly that all wave functions constructed from $\mathrm{SU}(n)_k$ WZW models can be written as projected Fermi sea states. In Sec.~\ref{sec:ph}, we derive the parent Hamiltonians for the $\mathrm{SU}(3)_k$ series. Numerical simulations of these wave functions are presented in Sec.~\ref{sec:Numerical}. This paper is summarized with an outlook in Sec.~\ref{sec:Summary}. Appendix~\ref{app:RepSU3} lists some useful identities for the SU(3) Lie algebra.

\section{Model wave functions}
\label{sec:MWF}

In this section, we discuss how to construct model wave functions from $\mathrm{SU}(n)_k$ WZW models. Section~\ref{sec:SUnk_WZW} briefly reviews the $\mathrm{SU}(n)_k$ WZW model and conformal block wave functions. In Sec.~\ref{sec:SUn1}, we discuss the wave functions constructed from $\mathrm{SU}(n)_1$ WZW models with an emphasis on $\mathrm{SU}(2)_1$ and $\mathrm{SU}(3)_1$ cases. We demonstrate that these wave functions can be formulated as projected Fermi sea states. In Sec.~\ref{sec:SUnk}, we generalize the results to $\mathrm{SU}(n)_k$ cases with $k \geq 2$.

\subsection{Brief review of the $\mathrm{SU}(n)_k$ WZW model and conformal block wave functions}
\label{sec:SUnk_WZW}

The WZW models are rational CFTs whose currents form a symmetry algebra called the affine Lie algebra. The $\mathrm{SU}(n)_k$ WZW model is defined by its (chiral) currents $J^a(z) = \sum_{n \in \mathbb{Z}} J_{n}^{a} z^{-n-1} \, (a=1,\ldots,n^2-1)$~\footnote{For our purpose, it is sufficient to consider the chiral part, so we leave out the antichiral part.}, where the modes satisfy the commutation relation
\begin{align}
[J^a_m,J^b_n]=if_{abc}J^c_{m+n}+k\delta_{ab}\delta_{m+n,0} \, .
\label{eq:KMalgebra}
\end{align}
Here $f_{abc}$ is the structure constant of the SU($n$) Lie algebra, and $k$ specifies the level of the WZW theory. The chiral primary fields of the $\mathrm{SU}(n)_k$ WZW model, denoted as $\phi_{\lambda}$, transform under SU($n$) irreducible representations (Irreps), where $\lambda$ is the label of the Irrep. For SU(2), the Irreps are distinguished by the ``spin'', i.e., $\lambda= 0, 1/2, \ldots$. For SU($n$) with $n \geq 3$, we use Young tableaux to label Irreps. In the $\mathrm{SU}(n)_k$ WZW model, each Young tableau with no more than $k$ columns and no more than $n-1$ rows has an associated primary field. For the $\mathrm{SU}(2)_k$ case, the primary fields correspond to SU(2) Irreps $\lambda= 0, 1/2, 1, \ldots, k/2$. Each primary field of the $\mathrm{SU}(n)_k$ WZW model has $d_{\lambda}$ components ($d_{\lambda}$: dimension of the Irrep $\lambda$); these components are written as $\phi^\alpha_{\lambda}$, where $\alpha=1,2,\ldots,d_{\lambda}$. As an example, the six primary fields of the SU(3)$_2$ WZW model correspond to the following Young tableaux:
\begin{align}
\underset{\mathbf{1}}{\LG{\bullet}} \quad  \underset{\mathbf{3}}{\young(~)} \quad \underset{\mathbf{\bar{3}}}{\young(~,~)} \quad \underset{\mathbf{6}}{\young(~~)} \quad \underset{\mathbf{\bar{6}}}{\young(~~,~~)} \quad \underset{\mathbf{8}}{\young(~~,~)} \, ,
\label{eq:SU3Irreps}
\end{align}
where a single dot $\bullet$ denotes the trivial (singlet) representation. The number below the Young tableau correspond to the dimension of the Irrep, and the bar indicates that it is the complex conjugate representation of another Irrep.

The construction of wave functions with CFT correlators was pioneered by Moore and Read in the study of FQH systems in the continuum~\cite{Moore1991}. This approach has also been adopted for lattice systems~\cite{Nielsen2011,Nielsen2012,TuHH2013b,TuHH2014a,ZhangHC2022,Cirac2010,Nielsen2012,TuHH2014b,TuHH2015,Hackenbroich2017}. We follow the latter approach below and construct lattice wave functions from the $\mathrm{SU}(n)_k$ WZW model.

Let us start with a lattice system with $N$ sites on the complex plane. The coordinates of the lattice sites are denoted by complex variables $z_i$ ($i=1,2,\ldots, N$). Each lattice site accommodates a spin that transforms under the $\mathrm{SU}(n)$ Irrep $\lambda$~\footnote{Here we restrict ourselves to the simplest case with identical $\mathrm{SU}(n)$ Irreps in each site. We note, however, that one might also consider (suitably chosen) different Irreps in different lattice sites.}, with local spin states denoted by $|\alpha\rangle$ ($\alpha = 1,2,\ldots,d_{\lambda}$). The many-body wave function is constructed as
\begin{align}
|\Phi\rangle =\sum_{\alpha_1,\ldots,\alpha_N}\Phi_{\alpha_1,\ldots,\alpha_N}(z_1,\ldots,z_N)|\alpha_1,\ldots,\alpha_N\rangle
\label{eq:mwf}
\end{align}
with the coefficients being conformal blocks
\begin{align}
\Phi_{\alpha_1,\ldots,\alpha_N}(z_1,\ldots,z_N) = \langle 0| \phi_{\lambda}^{\alpha_1}(z_1) \cdots \phi_{\lambda}^{\alpha_N}(z_N)|0\rangle,
\label{eq:mwf-1}
\end{align}
where $|0\rangle$ is the vacuum of the CFT. To ensure that Eq.~\eqref{eq:mwf-1} yields a single outcome, we further require that (i) the field $\phi_\lambda$ is a simple current (i.e., the fusion of two $\phi_\lambda$ has a unique outcome) and (ii) the fusion of $N$ primary fields $\phi_{\lambda}$ gives the trivial identity field (denoted as $\phi_0$ below). Such wave functions can be viewed as MPSs with an infinite-dimensional auxiliary Hilbert space (conformal Hilbert space) and are hence referred to as infinite-dimensional matrix product states (IDMPSs).

Although IDMPSs in Eq.~\eqref{eq:mwf-1} are formally well defined, the $\mathrm{SU}(n)_k$ conformal blocks are generally difficult to calculate,
except for $\mathrm{SU}(n)_1$ and $\mathrm{SU}(2)_2$ cases where free-field representations exist. This severely limits the study of these wave functions.
A key observation of the present work is that all $\mathrm{SU}(n)_k$ IDMPSs possess an \textit{exact} projected Fermi sea representation, valid even for finite-size systems.
Combining with recent algorithmic developments for converting (projected) Fermi sea states into MPSs~\cite{Fishman2015,WuYH2020,JinHK2020,Petrica2021,Aghaei2020,JinHK2022a,LiKL2025,LiuT2025a}, we can efficiently analyze the $\mathrm{SU}(n)_k$ IDMPSs using MPS techniques.

\subsection{$\mathrm{SU}(n)_1$ IDMPSs}
\label{sec:SUn1}

In this subsection, we show that $\mathrm{SU}(n)_1$ IDMPSs can be formulated as projected Fermi sea states.
The construction of $\mathrm{SU}(n)_1$ IDMPSs has been reported in Refs.~\cite{TuHH2014a,Bondesan2014}.
As the $\mathrm{SU}(n)_1$ WZW models have free boson representations, $\mathrm{SU}(n)_1$ IDMPSs can be explicitly written as correlators of chiral vertex operators of free bosons, which become Jastrow wave functions in the spin basis. We find that these Jastrow wave functions can be rewritten as projected Fermi sea states. To illustrate this, we start with two examples, $\mathrm{SU}(2)_1$ and $\mathrm{SU}(3)_1$ IDMPSs, and the generalization to $\mathrm{SU}(n)_1$ cases with $n\geq 4$ is straightforward.

\subsubsection{$n=2$ case}

The $\mathrm{SU}(2)_1$ WZW model has two primary fields, denoted as $\phi_0$ and $\phi_{1/2}$ with conformal weight $0$ and $1/4$, respectively. The fusion rules are
\begin{align}
\phi_0 \times \phi_0 = \phi_0, \quad \phi_0 \times \phi_{1/2} = \phi_{1/2}, \quad \phi_{1/2} \times \phi_{1/2}= \phi_0.
\end{align}
For constructing the $\mathrm{SU}(2)_1$ IDMPS~\cite{Cirac2010}, one should use $\phi_{1/2}$ in Eq.~\eqref{eq:mwf-1}, where $N$ must be even.
In the free boson representation, two components of the primary field $\phi_{1/2}$ are written as
\begin{align}
A_{\alpha}(z)=\kappa_{\alpha} :e^{i(3-2\alpha)\varphi(z)/\sqrt{2}}: \, , \quad (\alpha=1,2)
\label{eq:pf-su1a}
\end{align}
where $\varphi(z)$ is a chiral boson field and $: \cdots :$ represents normal ordering. $\kappa_{\alpha}$ are Klein factors that commute with vertex operators and
satisfy Majorana-like anticommutation relations
$\{\kappa_\alpha, \kappa_\beta\} = 2\delta_{\alpha\beta}$. The Klein factors ensure that the $\mathrm{SU}(2)_1$ IDMPS is an SU(2) singlet.

For the SU(2) case, it is more convenient to use $s=\frac{3-2\alpha}{2}$ ($s=\pm \frac{1}{2}$) to label local states.
Accordingly, the primary fields in Eq.~\eqref{eq:pf-su1a} are written as
\begin{align}
A_s(z)=\kappa_s :e^{is\sqrt{2}\varphi(z)}: \, .
\label{eq:pf_su21}
\end{align}
Substituting Eq.~\eqref{eq:pf_su21} into the chiral correlator in Eq.~\eqref{eq:mwf-1} gives the following Jastrow wave function:
\begin{align}
&\phantom{=} \;\; \Phi^{\mathrm{SU}(2)_1}_{s_1,\ldots,s_N}(z_1,\ldots,z_N) \nonumber \\
&=e^{i \frac{\pi}{2} \sum_{i:\mathrm{odd}}(2s_i-1)}\delta_{\sum_i s_i,0}\prod_{i<j}^N(z_i-z_j)^{2s_is_j}\nonumber\\
&=e^{i \frac{\pi}{2} \sum_{i:\mathrm{odd}}(2s_i-1)}\delta_{\sum_i s_i,0}\prod_{i<j}^{N}(z_i-z_j)^{-1/2}\nonumber\\
&\phantom{=} \times \prod_{i<j}^{s_i=s_j=1/2}(z_i-z_j) \prod_{i<j}^{s_i=s_j=-1/2}(z_i-z_j) \, ,
\label{eq:SU21IDMPS}
\end{align}
where the Marshall sign $e^{i \frac{\pi}{2} \sum_{i:\mathrm{odd}}(2s_i-1)}$ arises from permuting the Klein factors. The Jastrow product $\prod^N_{i<j} (z_i-z_j)^{-1/2}$ in Eq.~\eqref{eq:SU21IDMPS} is a state-independent overall factor and will be dropped. The Jastrow factor for each spin component, $\prod_{i<j}^{s_i=s_j= \pm 1/2}(z_i-z_j)$, takes the same form as that of the integer quantum Hall (IQH) wave function at unit filling.

If we use a fermionic representation for the spin-1/2 states, $|s_j\rangle = c^{\dagger}_{j,s}|0\rangle$, the following Fermi sea state (for each spin component $s$) generates the IQH Jastrow factor in Eq.~\eqref{eq:SU21IDMPS}:
\begin{align}
|\Psi_s\rangle=\prod^{N/2}_{p=1}d^{\dagger}_{p,s}|0\rangle
\label{eq:occupied_modes}
\end{align}
with (generally non-orthogonal) occupied single-particle orbitals
\begin{align}
d^{\dagger}_{p,s} = \sum_{i=1}^N (z_i)^{p-1}c^{\dagger}_{i,s} \, .
\label{eq:occupied_modes-1}
\end{align}
It is then straightforward to show that Eq.~\eqref{eq:SU21IDMPS} can be written as the following projected Fermi sea state:
\begin{align}
|\Phi^{\mathrm{SU}(2)_1}\rangle &\propto P_{\mathrm{G}} (|\Psi_{1/2}\rangle\otimes |\Psi_{-1/2}\rangle)  \nonumber \\
&= P_{\mathrm{G}} \prod_{s=\pm 1/2 } \prod^{N/2}_{p=1}d^{\dagger}_{p,s}|0\rangle \, ,
\label{eq:SU21PFS}
\end{align}
where $P_{\mathrm{G}}$ is the Gutzwiller projector imposing the single-occupancy constraint $\sum_{s=\pm 1/2} c^{\dagger}_{i,s} c_{i,s} = 1$ at each site.
The Marshall sign in Eq.~\eqref{eq:SU21IDMPS} appears naturally after using the mode expansion~\eqref{eq:occupied_modes-1} and permuting the fermionic modes into a site-ordered fashion. This proves the exact equivalence between the $\mathrm{SU}(2)_1$ IDMPS [Eq.~\eqref{eq:SU21IDMPS}] and the projected Fermi sea state [Eq.~\eqref{eq:SU21PFS}].

\subsubsection{$n=3$ case}

The $\mathrm{SU}(3)_1$ WZW model has three primary fields, denoted as $\phi_{0}$, $\phi_{\SM{\young(~)}}$, and $\phi_{\SM{\young(~,~)}}$ with conformal weight $0$, $1/3$, and $1/3$, respectively. Their fusion rules are given by
\begin{align}
&\phi_0 \times \phi_0 = \phi_0, \quad \phi_0 \times \phi_{\SM{\young(~)}} = \phi_{\SM{\young(~)}}, \quad \phi_{0} \times \phi_{\SM{\young(~,~)}}= \phi_{\SM{\young(~,~)}}, \nonumber \\
&\phi_{\SM{\young(~)}} \times \phi_{\SM{\young(~)}}=
\phi_{\SM{\young(~,~)}}, \quad \phi_{\SM{\young(~)}} \times \phi_{\SM{\young(~,~)}}=
\phi_{0}, \quad \phi_{\SM{\young(~,~)}} \times \phi_{\SM{\young(~,~)}}=
\phi_{\SM{\young(~)}}.
\end{align}
For constructing the $\mathrm{SU}(3)_1$ IDMPS~\cite{TuHH2014a,Bondesan2014}, we use $\phi_{\SM{\young(~)}}$ in Eq.~\eqref{eq:mwf-1} (choosing $\phi_{\SM{\young(~,~)}}$ gives the complex conjugate), where $N$ should be a multiple of three such that the fusion outcome of the fields in the IDMPS is $\phi_0$ and the wave function does not vanish. In the free boson representation, these primary fields read
\begin{align}
A_{\alpha}(z)= \kappa_{\alpha} :e^{i\sqrt{2}\vec{w}_{\alpha}\cdot\vec{\varphi}(z)}: \, ,   \quad (\alpha=1,2,3)
\label{eq:su_31}
\end{align}
where $\vec{\varphi}=(\varphi_1,\varphi_2)$ is a two-component chiral boson field and $\vec{w}_{\alpha}$ are weight vectors of local states $|\alpha\rangle$ in the SU(3) fundamental representation
\begin{align}
\vec{w}_1=(\frac{1}{2},\frac{1}{2\sqrt{3}}),\quad \vec{w}_2=(0,-\frac{1}{\sqrt{3}}),\quad \vec{w}_3=(-\frac{1}{2},\frac{1}{2\sqrt{3}}) \, .
\end{align}
Substituting Eq.~\eqref{eq:su_31} into Eq.~\eqref{eq:mwf-1}, we obtain the following Jastrow wave function:
\begin{align}
&\Phi^{\mathrm{SU}(3)_1}_{\alpha_1,\ldots,\alpha_N}(z_1,\ldots,z_N)= \mathrm{sgn}_3[\alpha_1,\ldots,\alpha_N]  \nonumber \\ 
&\phantom{=}  \times \delta_{\sum_i \vec{w}_{\alpha_i},\vec{0}}\prod^{N}_{i<j}(z_i-z_j)^{-1/3}\prod^{\alpha_i=\alpha_j}_{i<j}(z_i-z_j) \, ,
\label{eq:SU31IDMPS}
\end{align}
where $\mathrm{sgn}_3[s_1,\ldots,s_N]$ is a sign factor coming from the permutation of Klein factors~\cite{TuHH2014a,Bondesan2014} and the overall factor $\prod^N_{i<j} (z_i-z_j)^{-1/3}$ can be dropped.

Similar to the spin-1/2 case, we introduce a fermionic representation for the spin states in the SU(3) fundamental representation, $|\alpha_j\rangle = c^{\dagger}_{j,\alpha}|0\rangle$. Then, Eq.~\eqref{eq:SU31IDMPS} can be expressed as a projected Fermi sea state
\begin{align}
|\Phi^{\mathrm{SU}(3)_1}\rangle &\propto P_{\mathrm{G}} (|\Psi_{1}\rangle \otimes |\Psi_{2}\rangle \otimes |\Psi_{3}\rangle) \nonumber \\
&= P_{\mathrm{G}} \prod_{\alpha=1}^{3} \prod^{N/3}_{p=1}d^{\dagger}_{p,\alpha}|0\rangle \, ,
\label{eq:SU31PFS}
\end{align}
where $d^{\dagger}_{p,\alpha}$ takes the same form as Eq.~\eqref{eq:occupied_modes-1} (with spin index $s$ replaced by color index $\alpha$), $|\Psi_{\alpha}\rangle$ is the $1/3$-filled Fermi sea for color component $\alpha$, and the Gutzwiller projector $P_{\mathrm{G}}$ imposes the single-occupancy constraint $\sum_{\alpha= 1}^{3} c^{\dagger}_{i,\alpha} c_{i,\alpha} = 1$ for local states in the SU(3) fundamental representation. This establishes the exact equivalence between the $\mathrm{SU}(3)_1$ IDMPS [Eq.~\eqref{eq:SU31IDMPS}] and the projected Fermi sea state [Eq.~\eqref{eq:SU31PFS}].

\subsubsection{General cases}

After proving the equivalence between the $\mathrm{SU}(n)_1$ IDMPSs and the project Fermi sea states for $n=2$ and $3$, we briefly comment on how to extend the proof to all $n \geq 4$ cases. The $\mathrm{SU}(n)_1$ WZW model has $n$ primary fields corresponding to Young tableaux with a single column and $r=0,1,\ldots,n-1$ rows. We use the primary field $\phi_{\SM{\young(~)}}$ in the SU($n$) fundamental representation to construct the IDMPS~\cite{TuHH2014a,Bondesan2014}. As fusing $n$ $\phi_{\SM{\young(~)}}$ fields gives the identity, the number of sites, $N$, should be a multiple of $n$. In the free boson representation, the primary fields in the SU($n$) fundamental representation read
\begin{align}
A_{\alpha}(z)= \kappa_{\alpha} :e^{i\sqrt{2}\vec{w}_{\alpha}\cdot\vec{\varphi}(z)}: \, ,
\label{eq:su_n1}
\end{align}
where $\vec{\varphi}=(\varphi_1,\ldots,\varphi_{n-1})$ is a $(n-1)$-component chiral boson field and $\vec{w}_{\alpha}$ are weight vectors of local states $|\alpha\rangle$ in the SU($n$) fundamental representation
\begin{align}
\vec{w}_1&=(\frac{1}{2},\frac{1}{2\sqrt{3}},\ldots,\frac{1}{\sqrt{2n(n-1)}}) \, , \nonumber \\
\vec{w}_2&=(-\frac{1}{2},\frac{1}{2\sqrt{3}},\ldots,\frac{1}{\sqrt{2n(n-1)}}) \, , \nonumber \\
\vec{w}_3&=(0,-\frac{1}{2\sqrt{3}},\ldots,\frac{1}{\sqrt{2n(n-1)}})\, , \nonumber \\
&\vdots\nonumber \\
\vec{w}_n&=(0,0,\ldots,-\frac{n-1}{\sqrt{2n(n-1)}}) \, .
\end{align}

Substituting Eq.~\eqref{eq:su_n1} into Eq.~\eqref{eq:mwf-1}, we obtain
\begin{align}
&\Phi^{\mathrm{SU}(n)_1}_{\alpha_1,\ldots,\alpha_N}(z_1,\ldots,z_N)= \mathrm{sgn}_n[\alpha_1,\ldots,\alpha_N]  \nonumber \\ 
&\phantom{=}  \times \delta_{\sum_i \vec{w}_{\alpha_i},\vec{0}}\prod^{N}_{i<j}(z_i-z_j)^{-\frac{1}{n}}\prod^{\alpha_i=\alpha_j}_{i<j}(z_i-z_j) \, ,
\end{align}
where $\mathrm{sgn}_n[\alpha_1,\ldots,\alpha_N]$ is the sign factor from permuting $n$ different kinds of Klein factors. Dropping the overall factor, $\mathrm{SU}(n)_1$ IDMPSs are rewritten as projected Fermi sea states
\begin{align}
|\Phi^{\mathrm{SU}(n)_1}\rangle &\propto P_{\mathrm{G}} \prod_{\alpha=1}^{n} \prod^{N/n}_{p=1}d^{\dagger}_{p,\alpha}|0\rangle \, ,
\label{eq:SUn1PFS}
\end{align}
where the Fermi sea for each component is $1/n$-filled and the Gutzwiller projector $P_{\mathrm{G}}$ ensures the single-occupancy constraint $\sum_{\alpha=1}^{n} c^{\dagger}_{i,\alpha} c_{i,\alpha} = 1$ for local spin states in the SU($n$) fundamental representation.

\subsection{$\mathrm{SU}(n)_k$ IDMPSs with $k\geq 2$}
\label{sec:SUnk}

The $\mathrm{SU}(n)_k$ WZW models with $k\geq 2$ generally do not have a free-field representation, making it difficult to write down an explicit expression for IDMPSs in Eq.~\eqref{eq:mwf-1}.
The $\mathrm{SU}(2)_2$ WZW model is an exception as it has a free field representation in terms of three Majorana fermions (or one free boson and one Majorana fermion)~\cite{Nielsen2011,ZhangHC2021}.
For $\mathrm{SU}(2)_k$ WZW models, a key insight is that $\mathrm{SU}(2)_k$ ($k\geq 2$) IDMPSs can be obtained by symmetrizing $k$ copies of $\mathrm{SU}(2)_1$ IDMPSs~\cite{Quella2020}.
This provides the theoretical foundation for the symmetrization approach to $\mathrm{SU}(2)_k$ wave functions proposed in earlier works~\cite{Shastry1992,Narayan2004,Greiter2002,Greiter2009,Greiter-book,Greiter2014}. However, symmetrization itself does not provide a simple expression for the wave function, and it was still not possible to compute physical quantities for $\mathrm{SU}(2)_k$ ($k\geq 3$) IDMPSs. Nevertheless, the fact that the $\mathrm{SU}(2)_1$ IDMPS can be represented as a projected Fermi sea state implies that the $\mathrm{SU}(2)_k$ ($k\geq2$) IDMPSs are also projected Fermi sea states if we introduce an additional ``orbital'' index for fermions and view symmetrization as an extra local projection on top of the Gutzwiller projection. In this subsection, we start by reviewing the symmetrization procedure for $\mathrm{SU}(2)_k$ IDMPSs and then generalize it to $\mathrm{SU}(n)_k$ cases. We also demonstrate how to formulate $\mathrm{SU}(n)_k$ IDMPSs as projected Fermi sea states.

\subsubsection{$n = 2$ cases}

The $\mathrm{SU}(2)_k$ WZW model has $k+1$ primary fields denoted as $\phi_0, \phi_{1/2}, \ldots,  \phi_{k/2}$. Their fusion rules are given by
\begin{align}
\phi_l \times \phi_m = \phi_{|m-l|}+\phi_{|m-l|+1}+\cdots+\phi_{\mathrm{min}(m+l,k-m-l)}
\end{align}
with $l,m=0,1/2,\ldots,k/2$. The $\mathrm{SU}(2)_k$ IDMPS is constructed by using the simple current $\phi_{k/2}$~\cite{Nielsen2011} with a spin-$k/2$ at each site. We can write $\phi_{k/2}$ as the symmetrization 
\begin{align}
 B_s(z) = \sum_{s^1,s^2,\ldots,s^k} P^{s^1,s^2,\ldots,s^k}_s A^1_{s^1}(z) A^2_{s^2}(z),\ldots,A^k_{s^k}(z) \, ,
\label{eq:su_2k}
\end{align}
where $A^{\mu}_{s^{\mu}}(z)$ ($\mu=1,\ldots,k$) is the $\mu$-th copy of the $\mathrm{SU}(2)_1$ primary field $\phi_{1/2}$ given in Eq.~\eqref{eq:pf_su21}. The coefficients $P^{s^1,s^2,\ldots,s^k}_s$ are Clebsch–Gordan (CG) coefficients for symmetrizing $k$ spin-1/2's into a single spin-$k/2$, $|k/2;s\rangle = \sum_{s^1,s^2,\ldots,s^k=\pm 1/2} P^{s^1,s^2,\ldots,s^k}_s |1/2; s^1\rangle \otimes |1/2; s^2\rangle \otimes \cdots \otimes |1/2; s^k\rangle$ with $s=-k/2,-k/2+1,\ldots,k/2$. For instance, the (nonvanishing) CG coefficients for symmetrizing two spin-1/2's into a spin-1 are given by $P^{1/2,1/2}_{1}=P^{-1/2,-1/2}_{-1}=1$ and $P^{1/2,-1/2}_{0}=P^{-1/2,1/2}_{0}=1/\sqrt{2}$.

Substituting Eq.~\eqref{eq:su_2k} into Eq.~\eqref{eq:mwf-1} , the $\mathrm{SU}(2)_k$ IDMPSs are expressed as 
\begin{align}
&\phantom{=} \;\; \Phi^{\mathrm{SU}(2)_k}_{s_1,\ldots,s_N} (z_1,\ldots,z_N) 
 \nonumber \\
&=\langle0|B_{s_1}(z_1)\ldots B_{s_N}(z_N)|0\rangle \nonumber \\
&= \prod_{i=1}^N P^{s^1_i,s^2_i,\ldots,s^k_i}_{s_i} \prod_{\mu=1}^k \langle 0 |A_{s^{\mu}_1}(z_1)\ldots A_{s^{\mu}_N}(z_N)|0\rangle \nonumber \\
&= \prod_{i=1}^N P^{s^1_i,s^2_i,\ldots,s^k_i}_{s_i}\prod_{\mu=1}^k \Phi^{\mathrm{SU}(2)_1}_{s^{\mu}_1,s^{\mu}_2,\ldots,s^{\mu}_N}(z_1,\ldots,z_N) \, .
\label{eq:wf_su2k}
\end{align}
This shows that $\mathrm{SU}(2)_k$ IDMPSs can indeed be viewed as a symmetrization of $k$ copies of $\mathrm{SU}(2)_1$ IDMPSs~\cite{Quella2020}. In the discussion below, we use a short-hand notation
\begin{align}
\Phi^{\mathrm{SU}(2)_k}=\mathrm{Symm}(\overbrace{\Phi^{\mathrm{SU}(2)_1},\ldots, \Phi^{\mathrm{SU}(2)_1}}^{k})
\end{align}
for such symmetrized wave functions, where ``Symm'' stands for symmetrization.

The wave functions in Eq.~\eqref{eq:wf_su2k} representing $\mathrm{SU}(2)_k$ IDMPSs can be rewritten as projected Fermi sea states 
\begin{align}
|\Phi^{\mathrm{SU}(2)_k}\rangle \propto P_{\mathrm{Symm}} \prod_{\tau=1}^{k}\prod_{s=\pm 1/2} \prod^{N/2}_{p=1}d^{\dagger}_{p,\tau,s}|0\rangle \, ,
\label{eq:SU2kPFS}
\end{align}
where $\tau=1,\ldots,k$ is an orbital index. The fermions at each site must be singly occupied in each orbital, {$\sum_{s=\pm 1/2} c^{\dagger}_{i,\tau,s} c_{i,\tau,s} = 1 \; \forall \tau$}, so that they can reach spin-$k/2$ by symmetrization. We use $P_{\mathrm{Symm}}$ in Eq.~\eqref{eq:SU2kPFS} to implement symmetrization and projection onto spin-$k/2$ at each site (note that the single occupancy condition for each orbital is already implied).

\subsubsection{$n \geq 3$ cases}

We now use the symmetrization approach to construct $\mathrm{SU}(n)_k$ IDMPSs. The simple current of the $\mathrm{SU}(n)_k$ WZW model used for constructing the $\mathrm{SU}(n)_k$ IDMPSs is the primary field corresponding to the Young tableau with a single row and $k$ columns, which can be viewed as a symmetrization of $k$ $\mathrm{SU}(n)_1$ primary fields [Eq.~\eqref{eq:su_n1}] in the SU($n$) fundamental representation. Namely, the CG coefficients taking care of the symmetrization correspond to the tensor product decomposition of $k$ SU($n$) Irreps
\begin{align}
|\overbrace{\sm{\young(~~)}\cdots\sm{\young(~)}}^{k};\beta\rangle = \sum_{\alpha^1,\ldots,\alpha^k} P_{\beta}^{\alpha^1,\ldots,\alpha^k} |\sm{\young(~)};\alpha^1\rangle \otimes \cdots \otimes |\sm{\young(~)};\alpha^k\rangle \, ,
\end{align}
where $\beta$ labels basis states in the SU($n$) Irrep with $k$ horizontal boxes in the Young tableau. Thus, $\mathrm{SU}(n)_k$ IDMPSs can be written as 
\begin{align}
\Phi^{\mathrm{SU}(n)_k}=\mathrm{Symm}&(\overbrace{\Phi^{\mathrm{SU}(n)_1},\ldots, \Phi^{\mathrm{SU}(n)_1}}^{k}) \, .
\end{align}
This gives a systematic construction of all $\mathrm{SU}(n)_k$ IDMPSs.

Generalizing the projected Fermi sea formulation [Eq.~\eqref{eq:SU2kPFS}] for $\mathrm{SU}(2)_k$ IDMPSs, $\mathrm{SU}(n)_k$ IDMPSs have the following projected Fermi sea representation:

\begin{align}
|\Phi^{\mathrm{SU}(n)_k}\rangle \propto P_{\mathrm{Symm}} \prod_{\tau=1}^{k}\prod_{\alpha=1}^{n} \prod^{N/n}_{p=1}d^{\dagger}_{p,\tau,\alpha}|0\rangle \, ,
\label{eq:SUnkPFS}
\end{align}
where $P_{\mathrm{Symm}}$ implements local projections on each site, ensuring that for each site, i) every orbital is singly occupied and ii) $k$ fermions, each carrying an SU($n$) fundamental representation, are symmetrized into the SU($n$) Irrep corresponding to the Young tableau with $k$ horizontal boxes. The number of sites, $N$, must be a multiple of $n$ to define $|\Phi^{\mathrm{SU}(n)_k}\rangle$.

\section{Parent Hamiltonians}
\label{sec:ph}

In this section, we derive the parent Hamiltonians for $\mathrm{SU}(3)_k$ IDMPSs. The reason for concentrating on the SU(3) case is that $\mathrm{SU}(3)_k$ IDMPSs with $k \ge 2$ are lattice realizations of the so-called non-Abelian spin singlet states~\cite{Ardonne1999,Ardonne2001}, which form a particularly interesting family of FQH states supporting non-Abelian anyons. The analysis carried out in this section is generalizable to $\mathrm{SU}(n)_k$ cases with $n \ge 4$.

For deriving parent Hamiltonians of IDMPSs, a systematic approach was developed in Ref.~\cite{Nielsen2011}. This approach uses suitable CFT null vectors to construct a set of lattice operators annihilating the target IDMPS. Below we apply this method to derive the parent Hamiltonians for $\mathrm{SU}(3)_k$ IDMPSs.

Let us start with the SU(3)$_2$ case. For the SU(3)$_2$ WZW model, null vectors exist at Virasoro level $1$:
\begin{align}
|\lambda; \gamma\rangle = \sum_{a,\alpha} W_{a\alpha}^{\gamma} J^a_{-1} |\sm{\young(~~)};\alpha \rangle \, ,
\label{eq:nullvector}
\end{align}
where $J^a_{-1}$ ($a=1,\ldots,8$) are modes of the SU(3)$_2$ Kac-Moody currents [see Eq.~\eqref{eq:KMalgebra}], $|\sm{\young(~~)};\alpha\rangle$ ($\alpha=1,\ldots,6$) are chiral primary states transforming under the SU(3) Irrep \textbf{6} [see Eq.~\eqref{eq:SU3Irreps}], and $W_{a\alpha}^{\gamma}$ are coefficients that are determined by the null vector condition $\langle \lambda;\gamma| \lambda;\gamma \rangle=0$ ($\lambda$ denotes the SU(3) Irrep of the null vectors; see below). Actually, Eq.~\eqref{eq:nullvector} can be viewed as decomposing the tensor product of two SU(3) Irreps into a direct sum of Irreps:
\begin{align}
\underset{\mathbf{8}}{\sm{\young(~~,~)}} \otimes \underset{\mathbf{6}}{\sm{\young(~~)}}= \underset{{\textcolor{red}{\mathbf{24}}}}{\textcolor{red}{\sm{\young(~~~~,~)}}} \oplus \underset{\mathbf{\overline{15}}}{\sm{\young(~~~,~~)}} \oplus \underset{\mathbf{6}}{\sm{\young(~~)}} \oplus \underset{\mathbf{\bar{3}}}{\sm{\young(~,~)}} \, ,
\label{eq:tpd_su32}
\end{align}
where $J^{a}_{-1}$ and $|\sm{\young(~~)};\alpha\rangle$ transform under SU(3) Irreps \textbf{8} and \textbf{6}, respectively. We find that the null vectors belong to the SU(3) Irrep \textbf{24}, corresponding to the first Young tableau at the right-hand side of Eq.~\eqref{eq:tpd_su32} (marked in red). $W_{a\alpha}^{\gamma}$ are hence SU(3) CG coefficients fusing Irreps \textbf{8} and \textbf{6} into the Irrep \textbf{24}.

For the $\mathrm{SU}(3)_k$ WZW models in general ($k \geq 1$), the null vectors take the form
\begin{align}
&|\lambda; \gamma \rangle = \sum_{a,\alpha} W_{a\alpha}^{\gamma} J^a_{-1} |\overbrace{\sm{\young(~~)}\cdots\sm{\young(~)}}^{k};\alpha\rangle \, .
\label{eq:nullvector1}
\end{align}
The corresponding tensor product decomposition is
\begin{align}
&\phantom{=} \quad \underset{\mathbf{8}}{\sm{\young(~~,~)}} \otimes \underset{\mathbf{\frac{1}{2}(k+1)(k+2)}}{\overbrace{\sm{\young(~~)}\cdots\sm{\young(~)}}^{k}}  \nonumber \\
&= \underset{{\textcolor{red}{\mathbf{(k+2)(k+4)}}}}{\textcolor{red}{\overbrace{\sm{\young(~~,~)}\cdots{}^{\sm{\young(~)}}}^{k+2}}} \oplus \underset{\mathbf{\overline{\frac{3}{2}k(k+3)}}}{\overbrace{\sm{\young(~~~,~~)}\cdots{}^{\sm{\young(~)}}}^{k+1}} \oplus \underset{\mathbf{\frac{1}{2}(k+1)(k+2)}}{\overbrace{\sm{\young(~~)} \cdots\sm{\young(~)}}^{k}} \oplus \underset{\mathbf{k^2-1}}{\overbrace{\sm{\young(~~,~)}\cdots{}^{\sm{\young(~)}}}^{k-1}}\, .
\label{eq:tpd_su3k}
\end{align}
Note that the last Young tableau in Eq.~\eqref{eq:tpd_su3k} vanishes for $k=1$ and is better denoted as $\mathbf{\bar{3}}$ for $k=2$.

For constructing the parent Hamiltonians, there is no need to derive all CG coefficients $W_{a\alpha}^{\gamma}$ in Eq.~\eqref{eq:nullvector1}, and it suffices to have the projector $K_{a\alpha,b\beta} \equiv \sum_{\gamma} W_{a\alpha}^{\gamma} (W_{b\beta}^{\gamma})^{*}$~\cite{Nielsen2011}. Note that the orthonormality condition of the CG coefficients, $\sum_{a,\alpha} (W_{a\alpha}^{\gamma})^{*} W_{a\alpha}^{\gamma'} = \delta_{\gamma\gamma'}$, ensures that $K$ is a projector (i.e., $K^2 = K$). For computing $K$, we adapt a trick used in Ref.~\cite{Quella2020}. In terms of the Dynkin label, SU(3) Irreps are denoted as $(\lambda_1,\lambda_2)=(n_1 - n_2, n_2)$, where $n_{\nu}$ ($\nu=1,2$) is the number of boxes in the $\nu$-th row of the corresponding Young tableau. The dimension of the SU(3) Irrep $(\lambda_1,\lambda_2)$ is given by~\cite{Humphreys-Book}
\begin{align}
d_{(\lambda_1,\lambda_2)} = \frac{1}{2}(\lambda_1+1)(\lambda_2+1)(\lambda_1+\lambda_2+2) \, ,
\label{eq:su3dimension}
\end{align}
and its (quadratic) Casimir charge reads
\begin{align}
C^2_{(\lambda_1,\lambda_2)}=\frac{1}{3}(\lambda_1^2+\lambda_1\lambda_2+\lambda_2^2+3\lambda_1+3\lambda_2) \, ,
\label{eq:su3casimir}
\end{align}
Using the Dynkin labels, Eq.~\eqref{eq:tpd_su3k} is written as
\begin{align}
&\phantom{=} \;\;  (1,1) \otimes (k,0) \nonumber \\
&= \textcolor{red}{(k+1,1)} \oplus (k-1,2) \oplus (k,0)  \oplus (k-2,1) \, .
\label{eq:tpd_su3k-1}
\end{align}
As $K$ projects onto the subspace $(k+1,1)$ (and is hence SU(3)-symmetric), we can express it using the Casimir charges of the SU(3) Irreps involved in Eq.~\eqref{eq:tpd_su3k-1}:
\begin{align}
K &= \frac{C^2_{\mathrm{tot}}-C^2_{(k-1,2)}}{C^2_{(k+1,1)}-C^2_{(k-1,2)}}
\frac{C^2_{\mathrm{tot}}-C^2_{(k,0)}}{C^2_{(k+1,1)}-C^2_{(k,0)}} \nonumber \\
&\phantom{=} \; \times \frac{C^2_{\mathrm{tot}}-C^2_{(k-2,1)}}{C^2_{(k+1,1)}-C^2_{(k-2,1)}} \, ,
\label{eq:projector1}
\end{align}
where $C^2_{\mathrm{tot}}$ is the total Casimir operator living in the tensor product space $(1,1) \otimes (k,0)$. 
Denoting SU(3) generators in Irrep $(k,0)$ [Irrep $(1,1)$] as $t^a$ with matrix elements $t^{a}_{\alpha\beta}$ ($\mathfrak{t}^a$ with matrix elements $\mathfrak{t}^a_{bc} = -i f_{abc}$), the total Casimir operator $C^2_{\mathrm{tot}}$ is written as $C^2_{\mathrm{tot}} = \sum_{a=1}^{8}(\mathfrak{t}^a \otimes \mathbbm{1}_{(k,0)} + \mathbbm{1}_{(1,1)} \otimes t^{a})^2$, where $\mathbbm{1}$ with a subscript is the identity operator in the corresponding space. Substituting this into Eq.~\eqref{eq:projector1} and writing the matrix elements of the projector $K$ as $K^{ab}_{\alpha\beta} \equiv K_{a\alpha,b\beta}$ ($K^{ab}$ is an operator acting on the $(k,0)$ space), we obtain
\begin{align}
K^{ab}&=\frac{k+3}{3(k+1)}\delta_{ab} \mathbbm{1}_{(k,0)}-\frac{k+4}{(k+1)(k+3)}if_{abc}t^c   \nonumber \\
    &\phantom{=} \; -\frac{1}{(k+1)(k+3)}(t^a t^b+t^b t^a)+\frac{1}{k+1}g_{abc}t^c \, ,
\label{eq:su3koperator}
\end{align}
where $g_{abc}$ is the totally symmetric tensor of SU(3) [see Eq.~\eqref{eq:su3gtensor} in Appendix~\ref{app:RepSU3}]. In Eq.~\eqref{eq:su3koperator} and below, we adopt the Einstein summation convention (sum of the repeated indices is assumed). Further details leading to the derivation of Eq.~\eqref{eq:su3koperator} are given in Appendix~\ref{app:RepSU3}.

Using the null vector techniques developed in Ref.~\cite{Nielsen2011}, we obtain the following operators annihilating the $\mathrm{SU}(3)_k$ IDMPS:
\begin{align}
\Lambda^a_i=\sum_{j=1(\neq i)}^{N} w_{ij} K^{ab}_i t^b_j \, ,
\end{align}
where $i,j$ are site indices and $w_{ij}\equiv(z_i+z_j)/(z_i-z_j)$. Then, the positive semi-definite Hamiltonian $H^{\mathrm{SU}(3)_k} = \sum_{i=1}^{N} (\Lambda_i^a)^{\dagger} \Lambda_i^a$ has the $\mathrm{SU}(3)_k$ IDMPS as its zero-energy ground state. This parent Hamiltonian takes the following explicit form:
\begin{align}
H^{\mathrm{SU}(3)_k} &= \sum_{i \neq j} \frac{|w_{ij}|^2}{k+1} \left[\frac{k+6}{3}(\vec{t}_i \cdot \vec{t}_j) - \frac{2}{k+3}(\vec{t}_i \cdot \vec{t}_j)^2\right] \nonumber\\
&\phantom{=} \; +\sum_{i\neq j \neq k} \frac{w_{ij}^* w_{ik}}{k+1} \left[\frac{k+3}{3}(\vec{t}_j \cdot \vec{t}_k)  \right. \nonumber  \\
&\phantom{=} \; -\frac{k+4}{k+3} \, if_{abc}t^a_it^b_jt^c_k + g_{abc}t^a_it^b_jt^c_k  \nonumber  \\
&\phantom{=} \;  \left. -\frac{(\vec{t}_i \cdot \vec{t}_j)(\vec{t}_i \cdot \vec{t}_k)+(\vec{t}_i \cdot \vec{t}_k)(\vec{t}_i \cdot \vec{t}_j)}{k+3} \right] - E
\end{align}
with $E = -\frac{k(k+3)^2}{9(k+1)} \sum_{i \neq j} |w_{ij}|^2$ and $\vec{t}_i \cdot \vec{t}_j = t^a_i t^a_j$.

For a 1D periodic chain, the complex coordinates are chosen as $z_j=e^{i \frac{2\pi}{N} j}$ ($j = 1,\ldots,N$)~\cite{Cirac2010}, so that $N$ lattice sites are uniformly distributed on a unit circle. In this case, the parent Hamiltonian is simplified. We define the 1D parent Hamiltonian as $H^{\mathrm{SU}(3)_k}_{\mathrm{1D}} = \sum_{i=1}^{N} (\Lambda_i^a)^{\dagger} \Lambda_i^a + \frac{9+2N(k+3)}{6(k+1)}T^aT^a + \frac{1}{3(k+1)}g_{abc}T^aT^bT^c$, where $T^a=\sum_{i=1}^Nt^a_i$ is the total SU(3) spin operator. As $T^a |\Phi^{\mathrm{SU}(3)_k}\rangle = 0 \; \forall a$, total Casimir operators in the parent Hamiltonian only shift energy levels of excited states. The 1D parent Hamiltonian is explicitly written as
\begin{align}
H^{\mathrm{SU}(3)_k}_{\mathrm{1D}} &= \sum_{i\neq j} \frac{|w_{ij}|^2}{k+1} \left[ (k+4)(\vec{t}_i \cdot \vec{t}_j) - \frac{2}{k+3}(\vec{t}_i \cdot \vec{t}_j)^2 \right]  \nonumber \\
&\phantom{=} \; +\sum_{i\neq j \neq k}\frac{w_{i j}w_{i k}}{k+1} \frac{2(\vec{t}_i \cdot \vec{t}_j)(\vec{t}_i \cdot \vec{t}_k)}{k+3} - E_{\mathrm{1D}}
\end{align}
with $E_{\mathrm{1D}} = -\frac{k(k+3)^2}{27(k+1)}N(N^2+3)$.

Even though the parent Hamiltonians with long-range interactions are difficult to realize, they give us some hints on what types of interactions might be important to stabilize ground states that are in the same phase as the IDMPSs (see Refs.~\cite{Nielsen2013,Glasser2015,WuYH2016a,Nandy2019,ChenJY2021,Jaworowski2022,XuY2023} for various examples). This is particularly encouraging for the $\mathrm{SU}(3)_2$ case, as the $\mathrm{SU}(3)_2$ chiral spin liquid supports Fibonacci anyons which have been long sought due to their potential applications in fault-tolerant quantum information processing.

\section{Numerical results}
\label{sec:Numerical}

Section \ref{sec:MWF} demonstrates that all $\mathrm{SU}(n)_k$ IDMPSs have exact projected Fermi sea representations. 
Employing our newly developed algorithm that transforms projected Fermi sea states into MPSs~\cite{LiuT2025a}, 
we can obtain accurate MPS approximations for $\mathrm{SU}(n)_k$ IDMPSs, which enable us to investigate the physical properties of these wave functions. In this section, we present numerical results for the $\mathrm{SU}(2)_1$, $\mathrm{SU}(2)_2$, $\mathrm{SU}(3)_1$, and SU(3)$_2$ cases in both 1D and 2D lattices. For the 1D case, this construction yields quantum critical chains that fall into the universality class characterized by the corresponding WZW models. We validate the properties of the resulting chains by computing the central charge and the Klein bottle entropy. For the 2D case, we confirm that these IDMPSs represent $\mathrm{SU}(n)_k$ chiral spin liquids, for which we confirm by directly comparing the entanglement spectra~\cite{LiH2008} with CFT predictions.

\subsection{1D critical chains}
\label{sec:MPS1d}

When choosing the complex coordinates as $z_j = e^{i\frac{2\pi}{N}j}$ ($j=1,\ldots,N$), the $N$ lattice sites are uniformly placed on a unit circle in the complex plane. For the $\mathrm{SU}(n)_1$ IDMPS, it has been shown in Refs.~\cite{TuHH2014a,Quella2020} that the 1D parent Hamiltonian reduces to the $\mathrm{SU}(n)$ generalization~\cite{Kawakami1992,Ha1992} of the Haldane-Shastry model~\cite{Haldane1988a,Shastry1988}, whose low-energy effective theory is the $\mathrm{SU}(n)_1$ WZW model. We expect that the $\mathrm{SU}(n)_k$ IDMPSs are critical states described by the $\mathrm{SU}(n)_k$ WZW model, which would be in the same universality class as the $\mathrm{SU}(n)$ Andrei-Johannesson model~\cite{Andrei1984,Johannesson1986,Fuhringer2008}. In this subsection, we compute the central charge and the Klein bottle entropy from their MPS approximations to verify that these characteristic quantities are indeed in agreement with the expected universality class described by the corresponding WZW models.

\begin{figure}
    \centering
    \includegraphics[width=1.0\linewidth]{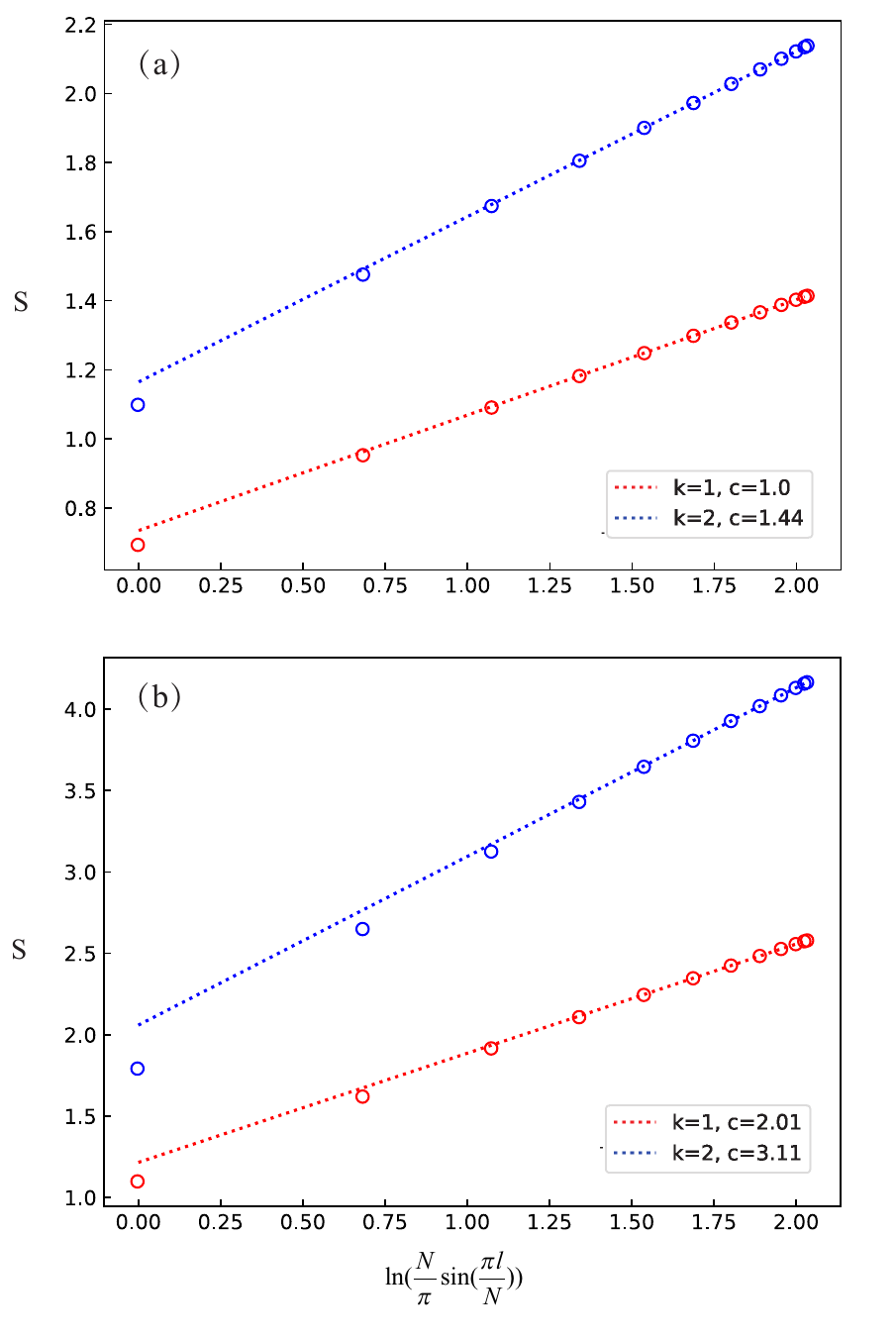}
    \caption{Entanglement entropy $S$ versus $\mathrm{ln}\left[\frac{N}{\pi}\mathrm{sin}(\frac{\pi l}{N})\right]$ for (a) $\mathrm{SU}(2)_{k=1,2}$ and (b) $\mathrm{SU}(3)_{k=1,2}$ IDMPSs on 1D periodic chains with $N=24$ sites. The central charges are obtained by linear fitting. When performing the linear fitting, the first five data points ($l=1,\ldots,5$) corresponding to small subsystem sizes have been discarded.} 
\label{fig:central charge}
\end{figure}

For the ground states of 1D periodic chains described by CFTs, the von Neumann entanglement entropy scales as~\cite{Holzhey1994,Vidal2003,Calabrese2004}
\begin{align}
S(l) = \frac{c}{3}\ln \left[\frac{N}{\pi }\sin\left(\frac{\pi l}{N} \right)\right] + c_2 \, ,
\end{align}
where $l$ is the subsystem size, $c$ is the central charge, and $c_2$ is a non-universal constant. As shown in Fig.~\ref{fig:central charge}, we extract the central charges of several IDMPSs from the entanglement entropy of periodic chains with length $N=24$. When converting IDMPSs to MPSs, we increase the bond dimension until the entanglement entropy is saturated. For $\mathrm{SU}(2)_1$, $\mathrm{SU}(2)_2$, and $\mathrm{SU}(3)_1$ IDMPSs, the necessary bond dimensions for achieving convergence are $D=600$, $2400$, $2000$, respectively. For the $\mathrm{SU}(3)_2$ IDMPS, the entanglement entropy is quite large even for $N=24$, so that we have to use bond dimension $D=12000$ for the MPS conversion. For the extracted central charge, a comparison between numerical results and theoretical values ($c= \frac{k(n^2-1)}{n+k}$ for the $\mathrm{SU}(n)_k$ WZW model~\cite{Francesco-Book}) is shown in Tab.~\ref{Tab:ck}. The numerically extracted central charges are quite close to the theoretical predictions. 

The Klein bottle entropy $K$ is another universal quantity for characterizing 2D CFTs~\cite{TuHH2017,TangW2017}. For rational CFTs, it is given by $K = \sum_a d_a/D$~\cite{TuHH2017}, where the sum is with respect to all primary fields, $d_a$ is the quantum dimension of the primary field $a$, and $D=\sqrt{\sum_a d_a^2}$ is the total quantum dimension. For $\mathrm{SU}(2)_k$ and $\mathrm{SU}(3)_k$ WZW models, the quantum dimensions can be read off from the explicit expression of the modular $S$ matrix~\cite{Gepner1986,Gannon1994}, which gives $K_{\mathrm{SU(2)}_k} = \sqrt{\frac{2}{k+2}}\cot\frac{\pi}{2(k+2)}$ and $K_{\mathrm{SU(3)}_k} = \sqrt{3}\cot\frac{\pi}{k+3}$.

The Klein bottle entropy $K$ can be computed from the squared overlap of the IDMPSs and the crosscap state, $K = |\langle \mathcal{C}_{\mathrm{latt}} |\Phi^{\mathrm{SU}(n)_k} \rangle|^2$~\cite{LiZQ2020,ZhangYS2023,TanBY2025}, with the crosscap state being defined as
\begin{align}
|\mathcal{C}_{\mathrm{latt}}\rangle = \prod_{j=1}^{N/2} \sum_{\alpha=1}^{d_{(k,0)}}|\alpha\rangle_j |\alpha\rangle_{j+N/2} \, .
\label{eq:crosscap}
\end{align}
Using the MPS approximations of the $\mathrm{SU}(n)_k$ IDMPSs, the crosscap state overlap can be computed easily~\cite{TanBY2025}. We have carried out this computation for various IDMPSs with $N=24$ sites. The comparison between numerical results and theoretical predictions is shown in Tab.~\ref{Tab:ck}, which again shows good agreement.

To further improve numerical accuracy of these results, it would be necessary to carry out simulations in larger systems and perform finite-size scaling analysis. For the $\mathrm{SU}(2)_1$ IDMPS, the crosscap state overlap was calculated analytically for finite-size systems~\cite{TanBY2025}, which deviates from the CFT prediction by a $1/N$ order correction. For all $\mathrm{SU}(n)_1$ IDMPSs, similar finite-size corrections are expected, as their parent Hamiltonians ($\mathrm{SU}(n)$ Haldane-Shastry models) are perfect lattice realizations of the $\mathrm{SU}(n)_1$ WZW CFTs (i.e., there are only irrelevant terms but no marginally irrelevant terms causing logarithmic corrections). For $\mathrm{SU}(n)_k$ IDMPSs with $k \geq 2$, marginally irrelevant terms are generally present in the low-energy effective theory associated with their parent Hamiltonians (see Refs.~\cite{Nielsen2011,Thomale2012} for numerical studies of the $\mathrm{SU}(2)_k$ cases). Using conformal perturbation theory, a $1/\ln(N)$ correction to the crosscap state overlap should be obtained~\cite{TanBY2025}.

\begin{table}[h]
\begin{tabular}{|c|cc|cc|}
\hline
      & \multicolumn{2}{c|}{numerical}       & \multicolumn{2}{c|}{theoretical}       \\ \hline
model & \multicolumn{1}{c|}{$c$}  & $K$  & \multicolumn{1}{c|}{$c$}  & $K$  \\ \hline
$\mathrm{SU}(2)_1$     & \multicolumn{1}{c|}{1.00}  & 1.401  & \multicolumn{1}{c|}{1}  & 1.414  \\ \hline
$\mathrm{SU}(2)_2$     & \multicolumn{1}{c|}{1.44}  & 1.612  & \multicolumn{1}{c|}{1.5}  & 1.707 \\ \hline
$\mathrm{SU}(3)_1$     & \multicolumn{1}{c|}{2.01}  & 1.688 & \multicolumn{1}{c|}{2} & 1.732 \\ \hline
$\mathrm{SU}(3)_2$     & \multicolumn{1}{c|}{3.11} & 2.203 & \multicolumn{1}{c|}{3.2} & 2.384\\ \hline
\end{tabular}
\caption{Comparison between numerical results and theoretical values for the central charge and the Klein bottle entropy. The calculations are performed on 1D chains with $N=24$ sites.}
\label{Tab:ck}
\end{table}

\subsection{2D chiral spin liquids}
\label{sec:MPS2d}

\begin{figure}
    \centering
    \includegraphics[width=0.8\linewidth]{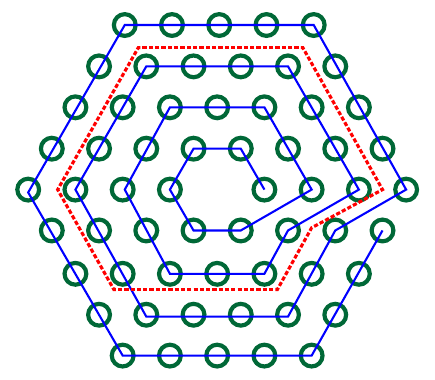}
    \caption{Lattice choice for the $\mathrm{SU}(n)_k$ IDMPSs in two dimensions. The green circles correspond to lattice sites, and the 1D path for defining the MPS is marked by a blue line. The red dashed line is the bipartite entanglement cut for computing the entanglement spectrum.}
\label{fig:disk}
\end{figure}

For the 2D case, we place the system on a disk with a radius of $R=4$ (with the central point excluded), as shown in Fig.~\ref{fig:disk}. This setup creates an annulus with $N=60$ sites. For the MPS conversion, we define a 1D path along the specified solid blue line, where a bipartite entanglement cut is performed at the center of the 1D chain, indicated by the red dashed line in Fig.~\ref{fig:disk}.

\begin{figure*}
    \centering
    \includegraphics[width=0.7\linewidth]{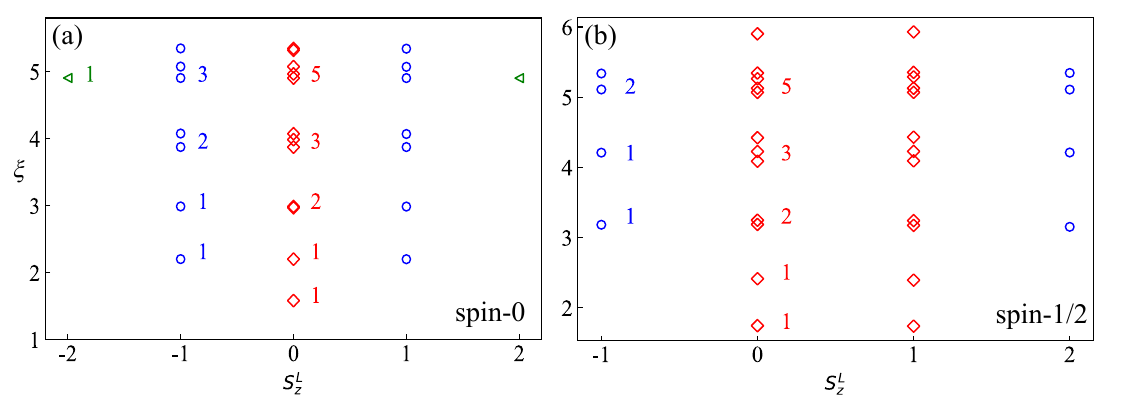}
    \caption{Entanglement spectra of the $\mathrm{SU}(2)_1$ IDMPSs (a) $\Phi^{\mathrm{SU}(2)_1}_{0}$ and (b) $\Phi^{\mathrm{SU}(2)_1}_{1/2}$. The entanglement levels are depicted versus the total $S^z$ of the subsystem (area surrounded by the red dashed line in Fig.~\ref{fig:disk}). The entanglement levels identified to be (quasi-)degenerate are marked using the same labels, and the numbers next to them denote the degeneracy. The data are obtained using MPS approximations with bond dimension $D=400$.}
\label{fig:SU(2)_1}
\end{figure*}

\begin{figure*}
    \centering
    \includegraphics[width=1.0\linewidth]{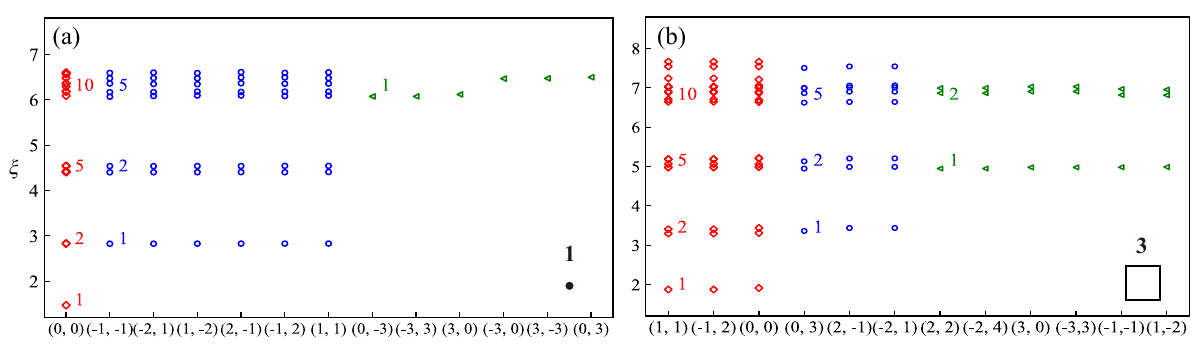}
    \caption{Entanglement spectra of the $\mathrm{SU}(3)_1$ IDMPSs (a) $\Phi^{\mathrm{SU}(3)_1}_{\mathbf{1}}$ and (b) $\Phi^{\mathrm{SU}(3)_1}_{\mathbf{3}}$. The entanglement levels are depicted versus the total weight (written as Dynkin labels) of the subsystem. The data are obtained using MPS approximations with bond dimension $D=480$.}
\label{fig:SU(3)_1}
\end{figure*}

With an MPS approximation in hand, the entanglement spectrum (i.e., the negative logarithm of the eigenvalues of the reduced density matrix) can be easily computed. The entanglement spectra of $|\Phi^{\mathrm{SU}(2)_1}\rangle$ and $|\Phi^{\mathrm{SU}(3)_1}\rangle$ are shown in Figs.~\ref{fig:SU(2)_1}(a) and \ref{fig:SU(3)_1}(a), respectively. For these two states, the counting of the low-lying entanglement spectra agrees with that of the identity sector of the corresponding chiral WZW model.

In the IDMPS setup, it is flexible to insert anyons~\cite{Nielsen2015}. Let us take the $\mathrm{SU}(2)_1$ IDMPS (discussed in Sec.~\ref{sec:SUn1}) as an example. We can insert a semion at the center of the disk by modifying the $\mathrm{SU}(2)_1$ IDMPS as
\begin{align}
&\phantom{=} \; \; \Phi^{\mathrm{SU}(2)_1}_{s_1,s_2,\ldots,s_N}(z_1,z_2,\ldots,z_N) \nonumber \\
&=\langle 0| A_{1/2}(z_0) A_{s_1}(z_1)\cdots A_{s_N}(z_{N}) A_{-1/2}(z_{N+1})|0\rangle \, ,
\label{eq:su21semion}
\end{align}
where the fields for constructing this IDMPS are already given in Eq.~\eqref{eq:pf-su1a}. Note that two extra operators are inserted at $z_0$ and $z_{N+1}$, where $z_0$ is the origin of the complex plane ($z_0 = 0$), and $z_{N+1}$ is far from the annulus (its specific value is not important). We denote the state defined via Eq.~\eqref{eq:su21semion} as $|\Phi^{\mathrm{SU}(2)_1}_{1/2}\rangle$ and the $\mathrm{SU}(2)_1$ IDMPS without extra operator insertions as $|\Phi^{\mathrm{SU}(2)_1}_{0}\rangle$ [see Eq.~\eqref{eq:SU21IDMPS}]. These two states are related by
\begin{align}
\Phi^{\mathrm{SU}(2)_1}_{1/2} = \Phi^{\mathrm{SU}(2)_1}_{0} \prod^{s_i=1/2}_i (z_0-z_i) \prod^{s_j=-1/2}_j(z_j-z_{N+1}) \, ,
\end{align}
where we have omitted spin and coordinate indices in $\Phi^{\mathrm{SU}(2)_1}_{1/2}$ and $\Phi^{\mathrm{SU}(2)_1}_{0}$ for simplicity. It indicates that $\Phi^{\mathrm{SU}(2)_1}_{1/2}$ can be constructed from $\Phi^{\mathrm{SU}(2)_1}_{0}$ by exciting one quasihole at the disk center for spin-up fermions and another quasihole outside the annulus for spin-down fermions. This type of wave function can also be written as a projected Fermi sea, just noting that both spin-up and spin-down parts still have the Slater determinant form. We corroborate this construction by calculating the entanglement spectrum of $\Phi^{\mathrm{SU}(2)_1}_{1/2}$ using its MPS approximation, and the result shown in Fig.~\ref{fig:SU(2)_1}(b) confirms that the counting of the low-lying entanglement spectra corresponds to that of the semion sector of the chiral $\mathrm{SU}(2)_1$ WZW model.

The insertion of anyons into the $\mathrm{SU}(3)_1$ IDMPS is done in a similar way. Starting from the wave function in Eq.~\eqref{eq:SU31IDMPS}, denoted as $\Phi^{\mathrm{SU}(3)_1}_{\mathbf{1}}$ below, the wave functions with anyons at the center of the disk transforming under SU(3) Irreps $\mathbf{3}$ and $\mathbf{\bar{3}}$ are given by
\begin{align}
\Phi^{\mathrm{SU}(3)_1}_{\mathbf{3}} &=\Phi^{\mathrm{SU}(3)_1}_{\mathbf{1}}\prod_i^{\alpha_i=1}(z_0-z_i)\prod_j^{\alpha_j=2,3}(z_j-z_{N+1}) \, , \nonumber \\
\Phi^{\mathrm{SU}(3)_1}_{\mathbf{\bar{3}}} &=\Phi^{\mathrm{SU}(3)_1}_{\mathbf{1}}\prod_i^{\alpha_i=1,2}(z_0-z_i)\prod_j^{\alpha_j=3}(z_j-z_{N+1}) \, ,
\end{align}
respectively. The entanglement spectrum of $\Phi^{\mathrm{SU}(3)_1}_{\mathbf{3}}$ is displayed in Fig.~\ref{fig:SU(3)_1}(b), and its counting agrees with the $\mathbf{3}$ sector of the chiral $\mathrm{SU}(3)_1$ WZW model. The entanglement spectrum of $\Phi^{\mathrm{SU}(3)_1}_{\mathbf{\bar{3}}}$ (not shown) is quite similar to that of $\Phi^{\mathrm{SU}(3)_1}_{\mathbf{3}}$.

\begin{figure*}
    \centering
    \includegraphics[width=1.0\linewidth]{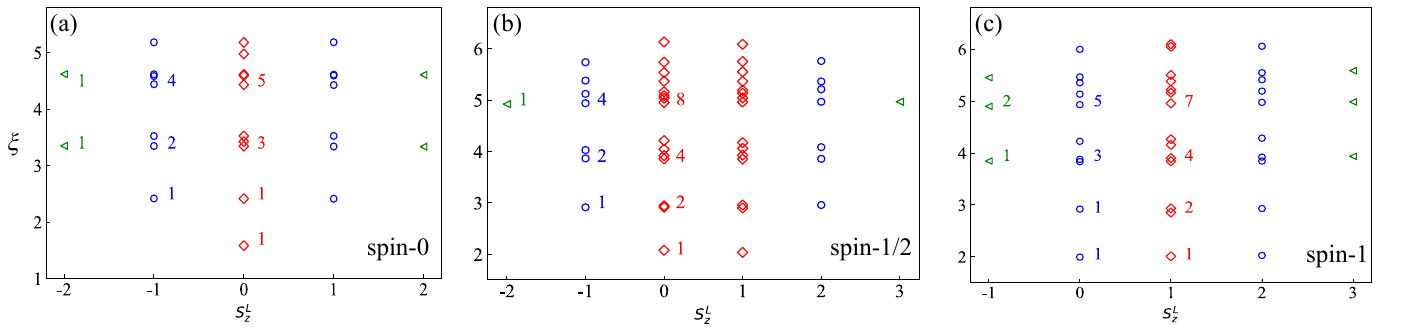}
    \caption{Entanglement spectra of the $\mathrm{SU}(2)_{2}$ IDMPSs (a) $\Phi^{\mathrm{SU}(2)_2}_{0}$, (b) $\Phi^{\mathrm{SU}(2)_1}_{1/2}$, and (c) $\Phi^{\mathrm{SU}(2)_2}_{1}$. The data are obtained using MPS approximations with bond dimension $D=480$.}
\label{fig:SU(2)_2}
\end{figure*}

\begin{figure*}
    \centering
    \includegraphics[width=0.8\linewidth]{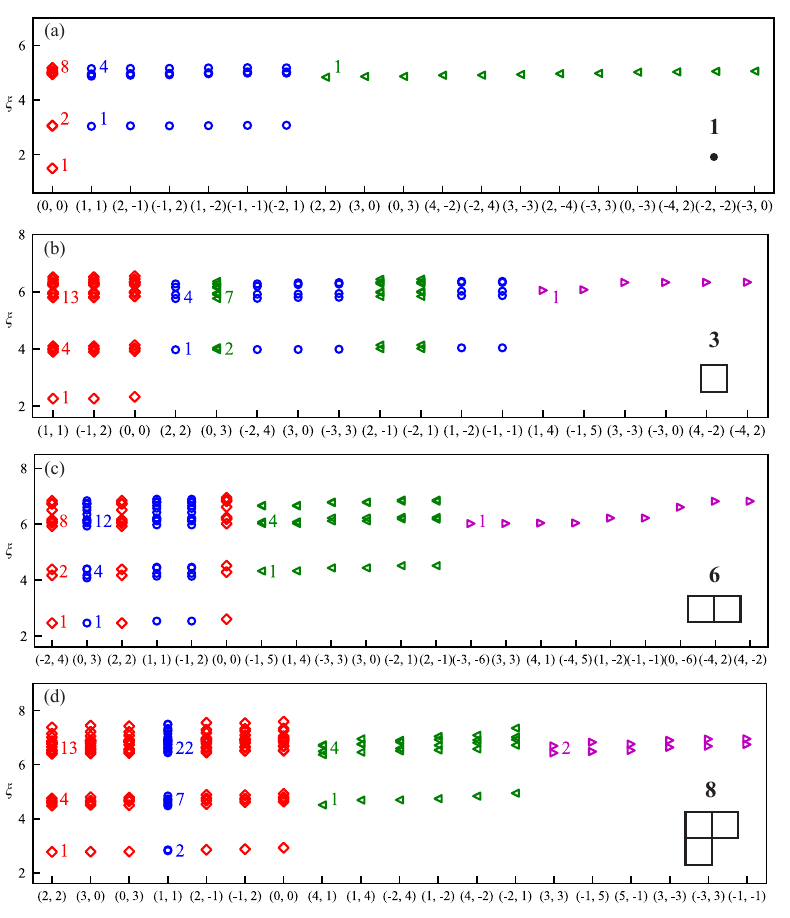}
    \caption{Entanglement spectra of the $\mathrm{SU}(3)_2$ IDMPSs (a) $\Phi^{\mathrm{SU}(3)_2}_{\mathbf{1}}$, (b) $\Phi^{\mathrm{SU}(3)_2}_{\mathbf{3}}$, (c) $\Phi^{\mathrm{SU}(3)_2}_{\mathbf{6}}$, and (d) $\Phi^{\mathrm{SU}(3)_2}_{\mathbf{8}}$. The data are obtained using MPS approximations with bond dimension $D=1200$.}
\label{fig:SU(3)_2}
\end{figure*}

Based on the observation in our previous work~\cite{LiuT2025a}, the anyonic sectors of $\mathrm{SU}(2)_{2}$ chiral spin liquids can be engineered by stacking and symmetrizing two layers of $\mathrm{SU}(2)_1$ chiral spin liquids with or without semions. Specifically, we expect the following combinations:
\begin{align}
\Phi^{\mathrm{SU}(2)_2}_0 &=\mathrm{Symm}(\Phi^{\mathrm{SU}(2)_1}_0,\Phi^{\mathrm{SU}(2)_1}_0) \, , \nonumber \\
\Phi^{\mathrm{SU}(2)_2}_{1/2} &=\mathrm{Symm}(\Phi^{\mathrm{SU}(2)_1}_0,\Phi^{\mathrm{SU}(2)_1}_{1/2}) \, , \nonumber\\
\Phi^{\mathrm{SU}(2)_2}_{1} &=\mathrm{Symm}(\Phi^{\mathrm{SU}(2)_1}_{1/2},\Phi^{\mathrm{SU}(2)_1}_{1/2}) \, .
\end{align}
The entanglement spectra for these three wave functions are shown in Figs.~\ref{fig:SU(2)_2}(a), (b) and (c), respectively. For the $\mathrm{SU}(3)_2$ IDMPS, we adopt the same strategy, which yields the following wave functions:
\begin{align}
\Phi^{\mathrm{SU}(3)_2}_{\mathbf{1}} &=\mathrm{Symm}(\Phi^{\mathrm{SU}(3)_1}_{\mathbf{1}},\Phi^{\mathrm{SU}(3)_1}_{\mathbf{1}}) \, , \nonumber \\
\Phi^{\mathrm{SU}(3)_2}_{\mathbf{3}} &=\mathrm{Symm}(\Phi^{\mathrm{SU}(3)_1}_{\mathbf{1}},\Phi^{\mathrm{SU}(3)_1}_{\mathbf{3}}) \, , \nonumber \\ 
\Phi^{\mathrm{SU}(3)_2}_{\mathbf{6}} &=\mathrm{Symm}(\Phi^{\mathrm{SU}(3)_1}_{\mathbf{3}},\Phi^{\mathrm{SU}(3)_1}_{\mathbf{3}}) \, , \nonumber \\
\Phi^{\mathrm{SU}(3)_2}_{\mathbf{8}} &=\mathrm{Symm}(\Phi^{\mathrm{SU}(3)_1}_{\mathbf{\bar{3}}},\Phi^{\mathrm{SU}(3)_1}_{\mathbf{3}}) \, , \nonumber \\
\Phi^{\mathrm{SU}(3)_2}_{\mathbf{\bar{3}}} &=\mathrm{Symm}(\Phi^{\mathrm{SU}(3)_1}_{\mathbf{1}},\Phi^{\mathrm{SU}(3)_1}_{\mathbf{\bar{3}}}) \, , \nonumber \\
\Phi^{\mathrm{SU}(3)_2}_{\mathbf{\bar{6}}} &=\mathrm{Symm}(\Phi^{\mathrm{SU}(3)_1}_{\mathbf{\bar{3}}},\Phi^{\mathrm{SU}(3)_1}_{\mathbf{\bar{3}}}) \, . 
\end{align}
We only depict the entanglement spectra for the first four wave functions in Fig.~\ref{fig:SU(3)_2}, as the entanglement spectra of $\Phi^{\mathrm{SU}(3)_2}_{\mathbf{\bar{3}}}$ and $\Phi^{\mathrm{SU}(3)_2}_{\SM{\mathbf{\bar{6}}}}$ are similar to those of $\Phi^{\mathrm{SU}(3)_2}_{\mathbf{3}}$ and $\Phi^{\mathrm{SU}(3)_2}_{\mathbf{6}}$, respectively. For $\mathrm{SU}(2)_2$ and $\mathrm{SU}(3)_2$ IDMPSs, the entanglement spectra for both identity and (nontrivial) anyonic sectors, shown in Figs.~\ref{fig:SU(2)_2} and \ref{fig:SU(3)_2}, agree very well with the respective sectors of the chiral WZW models, providing strong evidence that these IDMPSs represent chiral spin liquids with edge theories being the corresponding chiral WZW models.

\section{Summary and outlook}
\label{sec:Summary}

In summary, we have constructed model wave functions for spin systems using the $\mathrm{SU}(n)_k$ WZW models and accomplished a remarkable synergy of conformal field theory and parton approaches to non-Abelian states. Using the fermionic parton formulation, accurate MPS approximations can be found to facilitate the computation of many physical quantities. Based on an inspection of the central charge and the Klein bottle entropy, it is confirmed that the 1D IDMPSs describe critical systems whose low-energy physics is captured by the associated WZW models. For 2D systems with topological order, there are multiple topological sectors for which we can provide the wave functions with suitable anyon contents. Numerical results about the entanglement spectra in different sectors provide strong evidence that these wave functions represent $\mathrm{SU}(n)_k$ chiral spin liquids.

Since we have obtained the parent Hamiltonians for the $\mathrm{SU}(3)_{k}$ IDMPSs for all $k \geq 1$, it is natural to ask if this would help us to design more realistic short-range Hamiltonians that stabilizes the $\mathrm{SU}(3)_2$ chiral spin liquid. This would be an important step towards finding Fibonacci anyons in spin models. In view of the wide application of parton theory in strongly correlated systems, one may also try to reverse engineer some parton states to uncover their connection with CFT.

\section*{Acknowledgments}

We are grateful to Meng Cheng, Philippe Lecheminant, and Thomas Quella for helpful discussions. T.L. would like to thank Prof. Jan von Delft for hospitality and support during his visit at the Ludwig Maximilian University of Munich, where this work has been finalized. T.L. and T.X. are supported by the National Natural Science Foundation of China under Grant No. 12488201. Y.H.W. is supported by the National Natural Science Foundation of China under Grant No. 12174130.

\textbf{Data availability:}
The data that support the findings of this article are openly available~\cite{liu_zenodo}.

%\bibliography{refs}
%apsrev4-2.bst 2019-01-14 (MD) hand-edited version of apsrev4-1.bst
%Control: key (0)
%Control: author (8) initials jnrlst
%Control: editor formatted (1) identically to author
%Control: production of article title (-1) disabled
%Control: page (0) single
%Control: year (1) truncated
%Control: production of eprint (0) enabled
%

\appendix

\section{SU(3) Lie algebra and its representations}
\label{app:RepSU3}

In this Appendix, we briefly review some details about the SU(3) Lie algebra and its representations, which are useful for deriving some of the results in the main text.

\subsection{Fundamental representation}

The generators in the fundamental representation of the SU(3) Lie algebra can be chosen as
\begin{align}
\tau^1 & = \frac{1}{2} \begin{pmatrix}
0 & 1 & 0 \\
1 & 0 & 0 \\
0 & 0 & 0
\end{pmatrix} \ ,
\quad \tau^2=\frac{1}{2}\begin{pmatrix}
0 & -i & 0 \\
i & 0 & 0 \\
0 & 0 & 0
\end{pmatrix}, \nonumber \\
\tau^3 &= \frac{1}{2} \begin{pmatrix}
1 & 0 & 0 \\
0 & -1 & 0 \\
0 & 0 & 0
\end{pmatrix},
\quad \tau^4=\frac{1}{2}\begin{pmatrix}
0 & 0 & 1 \\
0 & 0 & 0 \\
1 & 0 & 0
\end{pmatrix}, \nonumber \\
\tau^5 & =\frac{1}{2}\begin{pmatrix}
0 & 0 & -i \\
0 & 0 & 0 \\
i & 0 & 0
\end{pmatrix}, 
\quad \tau^6=\frac{1}{2}\begin{pmatrix}
0 & 0 & 0 \\
0 & 0 & 1 \\
0 & 1 & 0
\end{pmatrix}, \nonumber \\
\tau^7 &=\frac{1}{2}\begin{pmatrix}
0 & 0 & 0 \\
0 & 0 & -i \\
0 & i & 0
\end{pmatrix}, 
\quad \tau^8=\frac{1}{2\sqrt{3}}\begin{pmatrix}
1 & 0 & 0 \\
0 & 1 & 0 \\
0 & 0 & -2
\end{pmatrix} ,
\end{align}
which are normalized as $\mathrm{tr}(\tau^a \tau^b)=\frac{1}{2}\delta_{ab}$. These generators satisfy the commutation relation of the SU(3) Lie algebra $[\tau^a,\tau^b]=if_{abc}\tau^c$, where the sum of the repeated index $c$ is assumed. We adopt such an Einstein summation convention in Appendix~\ref{app:RepSU3}. The structure constants $f_{abc}$ are totally antisymmetric, with the nonvanishing ones being $f_{123}=1$, $f_{147}=f_{246}=f_{257}=f_{345}=-f_{156}=-f_{367}=\frac{1}{2}$, $f_{458}=f_{678}=\frac{\sqrt{3}}{2}$ (othes are obtained by permutation).

For the fundamental representation of SU(3), we also have another useful equation: $\{\tau^a,\tau^b\}=\frac{1}{3}\delta_{ab}+g_{abc}\tau^c$,
where $g_{abc} = 2 \, \mathrm{tr}[\{\tau^a,\tau^b\}\tau^c]$ is a totally symmetric tensor. The non-vanishing elements of $g_{abc}$ are given by
\begin{align}
    g_{118}&=g_{228}=g_{338}=-g_{888}=\frac{1}{\sqrt{3}} \, , \nonumber \\
    g_{448}&=g_{558}=g_{668}=g_{778}=-\frac{1}{2\sqrt{3}} \, ,  \nonumber \\
    g_{443}&=g_{553}=g_{146}=g_{157}=g_{256}=\frac{1}{2} \, , \nonumber \\
    g_{663}&=g_{773}=g_{247}=-\frac{1}{2} \, ,
\label{eq:su3gtensor}
\end{align}
and the remaining ones are obtained by permutation.
It can be easily proven that $f_{abc}$ and $g_{abc}$ satisfy
\begin{align}
f_{cda}f_{cdb} =3\delta_{ab}, \quad 
g_{cda}g_{cdb} =\frac{5}{3}\delta_{ab} \, .
\end{align}

\subsection{Irreducible representation $(k,0)$}

For the SU(3) Irrep $(k,0)$ with $k \geq 2$, $\{t^a,t^b\}=\frac{1}{3}\delta_{ab}+g_{abc}t^c$ no longer holds. However, a totally symmetric tensor for the Irrep $(k,0)$ can be defined as
\begin{align}
    G_{abc}(k) \equiv 2 \, \mathrm{tr}[\{t^a,t^b\}t^c] \, ,
\end{align}
which is related to $g_{abc}$ by
\begin{align}
   G_{abc}(k) = g_{abc} \left(\frac{3k}{20}+\frac{3k^2}{8}+\frac{k^3}{3}+\frac{k^4}{8}+\frac{k^5}{60} \right) \, .
\label{eq:Gabc}
\end{align}
For $k=1$, $G_{abc}(k) = g_{abc}$.

For the SU(3) Irrep $(k,0)$ with $k \geq 1$, the following identities hold:
\begin{align}
\label{eq:tools1}
C^2_{(k,0)} &=t^at^a=\frac{k^2}{3}+k \, , \\
t^at^bt^a &=(C^2_{(k,0)}-\frac{3}{2})t^b \, , \\
t^a(t^bt^c+t^ct^b)t^a &= (C^2_{(k,0)} - 4)(t^bt^c+t^ct^b) \nonumber \\
&\phantom{=} \; + C^2_{(k,0)} \delta_{bc}+(k+\frac{3}{2})g_{bcd}t^d  \, , \\
-ig_{cah}f_{cde}t^ht^e &=-\frac{2k+3}{12}if_{adf}t^f-\frac{3}{4}g_{adf}t^f  \, , \\
\label{eq:tools5}
g_{abc}t^bt^c &= \frac{1}{6}\left(2k+3 \right) t^a \, ,
\end{align}
where $C^2_{(k,0)}$ is the Casimir charge for the SU(3) Irrep $(k,0)$.

Below we derive Eq.~\eqref{eq:su3koperator} in the main text. The total Casimir operator for the tensor product of SU(3) Irreps $(1,1) \otimes (k,0)$ is written as
\begin{align}
C^2_{\mathrm{tot}} &= \sum_{a=1}^{8}(\mathfrak{t}^a \otimes \mathbbm{1}_{(k,0)} + \mathbbm{1}_{(1,1)} \otimes t^{a})^2\nonumber \\
&= (C^2_{(1,1)}+C^2_{(k,0)})\mathbbm{1}_{(1,1)}\otimes\mathbbm{1}_{(k,0)}+2 \mathfrak{t}^c \otimes t^c \, ,
\end{align}
and its matrix elements are given by
\begin{align}
(C^2_{\mathrm{tot}})^{ab}_{\alpha\beta}=(C^2_{(1,1)}+C^2_{(k,0)})\delta_{ab}\delta_{\alpha\beta} -2if_{abc}t^{c}_{\alpha\beta} \, .
\label{eq:Casimir}
\end{align}
Substituting Eq.~\eqref{eq:Casimir} into Eq.~\eqref{eq:projector1}, we obtain
\begin{align}
K=\frac{(9+3k)+(30+8k)M+(28+4k)M^2+8M^3}{(1+k)(3+k)(3+2k)}
\label{eq:projector2}
\end{align}
with $M_{a\alpha,b\beta}=-if_{abc}t^c_{\alpha\beta}$. If we define $M_{a\alpha,b\beta} \equiv M^{ab}_{\alpha\beta}$ and treat $M^{ab}_{\alpha\beta}$ as an operator acting on the SU(3) Irrep $(k,0)$, we have
\begin{align}
(M^2)^{ab} &=(t^at^c-t^ct^a)(t^ct^b-t^bt^c) \nonumber \\
&=t^ct^ct^at^b+\frac{t^c(t^at^b+t^bt^a)t^c}{2}+i\frac{f_{abd}}{2}t^ct^dt^c \nonumber \\
&\phantom{=} \; -t^ct^at^ct^b-t^at^ct^bt^c\nonumber\\
&=\frac{1}{2}(\frac{k^2}{3}+k) \delta_{ab}-\frac{1}{2}(t^at^b+t^bt^a)+\frac{3}{4}if_{abd}t^d \nonumber \\
&\phantom{=} \; +(\frac{k}{2}+\frac{3}{4})g_{abd}t^d \, ,
\label{eq:M2}
\end{align}
and
\begin{align}
(M^3)^{ab}&= -(\frac{k^2}{6}+\frac{k}{2})if_{abd} t^d-\frac{1}{2}(t^at^c+t^ct^a)(t^bt^c-t^ct^b)\nonumber \\
&\phantom{=} \; -\frac{3}{4}(M^2)_{ab}-i\frac{(2k+3)g_{ach}f_{cbe}t^ht^e}{4} \nonumber\\
&=-\frac{5}{8}(\frac{k^2}{3}+k)\delta_{ab} +\frac{11}{8}(t^at^b+t^bt^a)\nonumber\\
&\phantom{=} \; -(\frac{k^2}{4}+\frac{3k}{4}+\frac{3}{8})if_{abd}t^d -(k+\frac{3}{2})g_{abd}t^d \, .
\label{eq:M3}
\end{align}
Substituting $M^2$ and $M^3$ into Eq.~\eqref{eq:projector2} gives the final form of the projector $K$ in Eq.~\eqref{eq:su3koperator}.

\end{document}